\documentclass{aa}  

\usepackage{graphicx}
\usepackage{txfonts}
\begin{document}

   \title{LOFAR observations of the quiet solar corona}


   \author{C. Vocks\inst{1}
     \and G. Mann\inst{1}
     \and F. Breitling\inst{1}
     \and M. M. Bisi\inst{2}
     \and B. D\k{a}browski\inst{3}
     \and R. Fallows\inst{4}
     \and P. T. Gallagher\inst{5}
     \and A. Krankowski\inst{3}
     \and J. Magdaleni\'{c}\inst{6}
     \and C. Marqu\'e\inst{6}
     \and D. Morosan\inst{5}
     \and H. Rucker\inst{7}
   }

   \institute{Leibniz-Institut f\"ur Astrophysik Potsdam,
              An der Sternwarte 16, 14482 Postdam, Germany\\
              \email{cvocks@aip.de}
   \and
   RAL Space, Science \& Technology Facilities Council — Rutherford
     Appleton Laboratory, Harwell Campus, Oxford, Oxfordshire OX11 0QX, UK
   \and
   Space Radio-Diagnostics Research Centre, University of Warmia
     and Mazury, Olsztyn, Poland
   \and
   ASTRON — the Netherlands Institute for Radio Astronomy, Postbus
     2, 7990 AA Dwingeloo, The Netherlands
   \and
   School of Physics, Trinity College Dublin, Dublin 2, Ireland
   \and
   Solar-Terrestrial Center of Excellence - SIDC, Royal
     Observatory of Belgium, Av. Circulaire 3, B-1180 Brussels, Belgium
   \and
   Commission for Astronomy, Austrian Academy of Sciences,
     Schmiedlstrasse 6, 8042, Graz, Austria
             }

   \date{Received ; accepted }

 
  \abstract
   {The quiet solar corona emits meter-wave thermal bremsstrahlung.
     Coronal radio emission can only propagate above that radius,
     $R_\omega$, where the local plasma frequency equals the observing
     frequency. The radio interferometer LOw Frequency ARray (LOFAR)
     observes in its low band (10 -- 90 MHz) solar radio emission
     originating from the middle and upper corona.}
   {We present the first solar aperture synthesis imaging observations in the
     low band of LOFAR in 12 frequencies each separated by 5 MHz. From each of
     these radio maps we infer $R_\omega$, and a scale height temperature,
     $T$. These results can be combined into coronal density and temperature
     profiles.}
   {We derived radial intensity profiles from the radio images. We focus
     on polar directions with simpler, radial magnetic field
     structure. Intensity profiles were modeled by ray-tracing simulations,
     following wave paths through the refractive solar corona, and including
     free-free emission and absorption. We fitted model profiles to
     observations with $R_\omega$ and $T$ as fitting parameters.}
   {In the low corona, $R_\omega < 1.5$ solar radii, we find
     high scale height temperatures up to $2.2\times 10^6$ K, much more
     than the brightness temperatures usually found there. But if all
     $R_\omega$ values are combined into a density profile, this profile can
     be fitted by a hydrostatic model with the same 
     temperature, thereby confirming this with two independent methods. The
     density profile deviates from the hydrostatic model above 1.5 solar
     radii, indicating the transition into the solar wind.}
   {These results demonstrate what information can be gleaned from
     solar low-frequency radio images. The scale height temperatures we find
     are not only higher than brightness temperatures, but also than
     temperatures derived from coronograph or extreme ultraviolet (EUV) data.
     Future observations will provide continuous frequency coverage. This
     continuous coverage eliminates the need for local hydrostatic density
     models in the data analysis and enables the analysis of more complex
     coronal structures such as those with closed magnetic fields.}


   \maketitle
%

\section{Introduction}

Radio observations of the Sun at frequencies below 100 MHz probe the middle
and upper corona, since radio waves cannot propagate through regions where
the wave frequency is lower than the local plasma frequency,
$\omega_\mathrm{p} = \sqrt{N_\mathrm{e}\,e^2 / (m_\mathrm{e}\,\epsilon_0)}$,
with electron number density, $N_\mathrm{e}$, elementary charge, $e$, electron
mass, $m_\mathrm{e}$, and vacuum permittivity, $\epsilon_0$. For 100 MHz, this
corresponds to a maximum density of $N_\mathrm{e} = 1.2\times
10^{14}\,\mathrm{m}^{-3}$, that is generally exceeded in the lower corona.
Interferometric imaging observations at low frequencies are well suited for
gaining insights into upper coronal structures. This technique has been used
over many decades.

\citet{wild70} observed the Sun with the Culgoora radio telescope at 80
MHz and determined flare positions, and also studied the quiet solar
radio emission and found brightness temperatures mostly below $10^6$
K. Culgoora data were also used by \citet{dulk77} for coronal hole
studies. These authors combined the data with microwave and extreme ultraviolet (EUV) observations to
constrain analytical transition region and coronal models.
\citet{kundu83} observed the Sun with the Clark Lake telescope between
100 MHz and 39 MHz, and found generally good agreement between radio maps and
white-light coronograph images \citep{kundu87}. \citet{wang87} used this
instrument to observe coronal holes and found brightness temperatures of not
more than $7\times 10^5$ K at the solar disk center at 73.8 MHz.
\citet{ramesh10} studied coronal streamers and tried to determine
magnetic fields in these streamers with the Gauribidanur Radioheliograph
\citep{ramesh98}. The Nan\c{c}ay Radioheliograph \citep[NRH; ][]{kerdraon97}
covers higher frequencies between 150 and 450 MHz, and has been used, for
example, by \citet{lantos99} to study medium- and large-scale features in the
quiet solar corona, such as coronal holes.

The interpretation of radio images of the solar corona requires modeling of
refractive effects, since the refractive index of a plasma approaches zero for
plasma wave frequencies near the local plasma frequency,
$\omega_\mathrm{p}$. \citet{alissandrakis94} calculated brightness
temperature maps for simple coronal hole models based on ray-tracing
calculations that consider refraction and opacity effects. The comparison of
the model results with NRH data leads to the conclusion that knowledge of coronal structures is important for the interpretation of radio maps. A
recent review by \citet{shibasaki11}, covering wavelength ranges from millimeter to
meter waves, also highlighted the importance of improved modeling of refraction
and scattering effects in the solar corona.

\citet{mercier15} collected NRH data over one solar cycle and derived
coronal density and temperature models in equatorial and polar
directions. These authors found brightness temperatures around $6\times 10^5$ K at the
lowest frequency of 150 MHz, but much higher scale height temperatures, around
$1.6\times 10^6$ K in the equatorial and up to $2.2\times 10^6$ K in the polar
direction.

The LOw Frequency ARray \citep[LOFAR;][]{vanhaarlem13} is a novel radio
interferometer that observes the sky in two frequency bands of 10 -- 90 MHz
(low band) and 110 -- 250 MHz (high band). The LOFAR telescope consists of a
central core of 24 antenna stations within a 3 km $\times$ 2 km area near
Exloo, the Netherlands, 14 remote stations spread over the Netherlands, and
international stations in France, Germany, Ireland, Poland, Sweden, and the
UK. The LOFAR baselines extend from tens of meters in the core up to 1550 km
across Europe. With this setup, LOFAR combines a high angular resolution below
1'' with a large collecting area and thus high sensitivity.

As a versatile radio telescope, it has uses in many fields of radio
astronomy. Science from LOFAR is organized around key science projects (KSPs). One
of the KSPs is Solar Physics and Space Weather with LOFAR \citep{mann11},
which studies all aspects of the active Sun and its effects on interplanetary
space, including the space environment of the Earth, which are commonly referred to
as space weather.

The frequency range of LOFAR covers the middle (high band) and upper (low band)
corona. However, scattering of radio waves on coronal turbulence prevents the
arc-second resolution provided by international baselines. \citet{mercier15b}
reported minimum source sizes of 31'' at 327 MHz and 35'' at 236 MHz. So, the
angular resolution of solar imaging with LOFAR is restricted to several tens
of arc-seconds at best. For LOFAR frequencies, this corresponds to
maximum baseline lengths of a few 10 km. Such baselines are provided by the
LOFAR core and the nearest remote stations.

Previous work on solar observations with LOFAR has focused on the active
Sun, for example, studies of type III radio bursts \citep{morosan14}. Here, we present
quiet-Sun observations in the low band on multiple frequencies between 25 and
79 MHz. A ray-tracing method is presented that is used to reproduce observed
intensity profiles with coronal density and scale height temperature as model
parameters. As the different frequencies probe different layers of the solar
corona, these results can be used to construct coronal density and temperature
profiles, at lower frequencies, i.e.,~larger heights, than most of the earlier
works presented above.

\section{LOFAR low-band images of the quiet Sun}
The appearance of the radio Sun can change within seconds, especially during
solar radio bursts. Therefore, solar imaging with LOFAR is normally snapshot
imaging with a short integration time of 1 s, or even 0.25 s, to
capture the dynamics of solar radio bursts. But for quiet-Sun studies, Earth
rotation aperture synthesis can be used to improve uv coverage, i.e.,~taking
advantage of the rotation of the Earth to increase the variety of different baseline
lengths and orientations of the interferometric array.

\subsection{External calibration with Tau A}
The structure of the radio Sun is a priori unknown and can vary substantially
with time. The thermal background emission from the $10^6$ K hot corona can
be estimated easily, but the intensity of solar radio bursts can exceed this
by up to four orders of magnitude \citep{mann10}. Nonthermal radio emission
from active regions, including radio bursts, can appear as compact or extended
sources with unknown intensities somewhere on the solar disk or even off the
limb. Measuring their position relative to the Sun is important for the
interpretation of the physical processes in the solar corona, since this measurement enables
an alignment with observations in other wavelength ranges such as optical,
EUV, or X-rays.

This precludes the use of a pure self-calibration approach, which is based on
an initial sky model with some prescribed solar intensity and center of
brightness. Therefore, an external calibration source is needed both for
amplitude and phase calibration. This needs to be a strong source to prevent
the active Sun from interfering with it easily. During the commissioning
phase of LOFAR, Tau A was found to be a suitable calibrator, which can be used
from late spring until the end of August. During the solar observations,
another LOFAR beam was pointed toward Tau A, recording calibration data on
the same frequencies as for the Sun.

However, one caveat needs to be taken into consideration for these
observations on 8 August: Tau A is located about 50 degree west of the Sun
at this time of the year, so it sinks toward the horizon during the observing
period of eight hours, centered around local noon. Hence its radio signal has to
traverse a longer path through the atmosphere, including the ionosphere. The
intensity of the Tau A signal decreases in the second half of the observing
period. Since Tau A is also used for amplitude calibration, a spurious
intensity increase would be transferred to the Sun.
\begin{figure*}[ht]
\hspace*{1.5cm}\includegraphics[width=6cm]{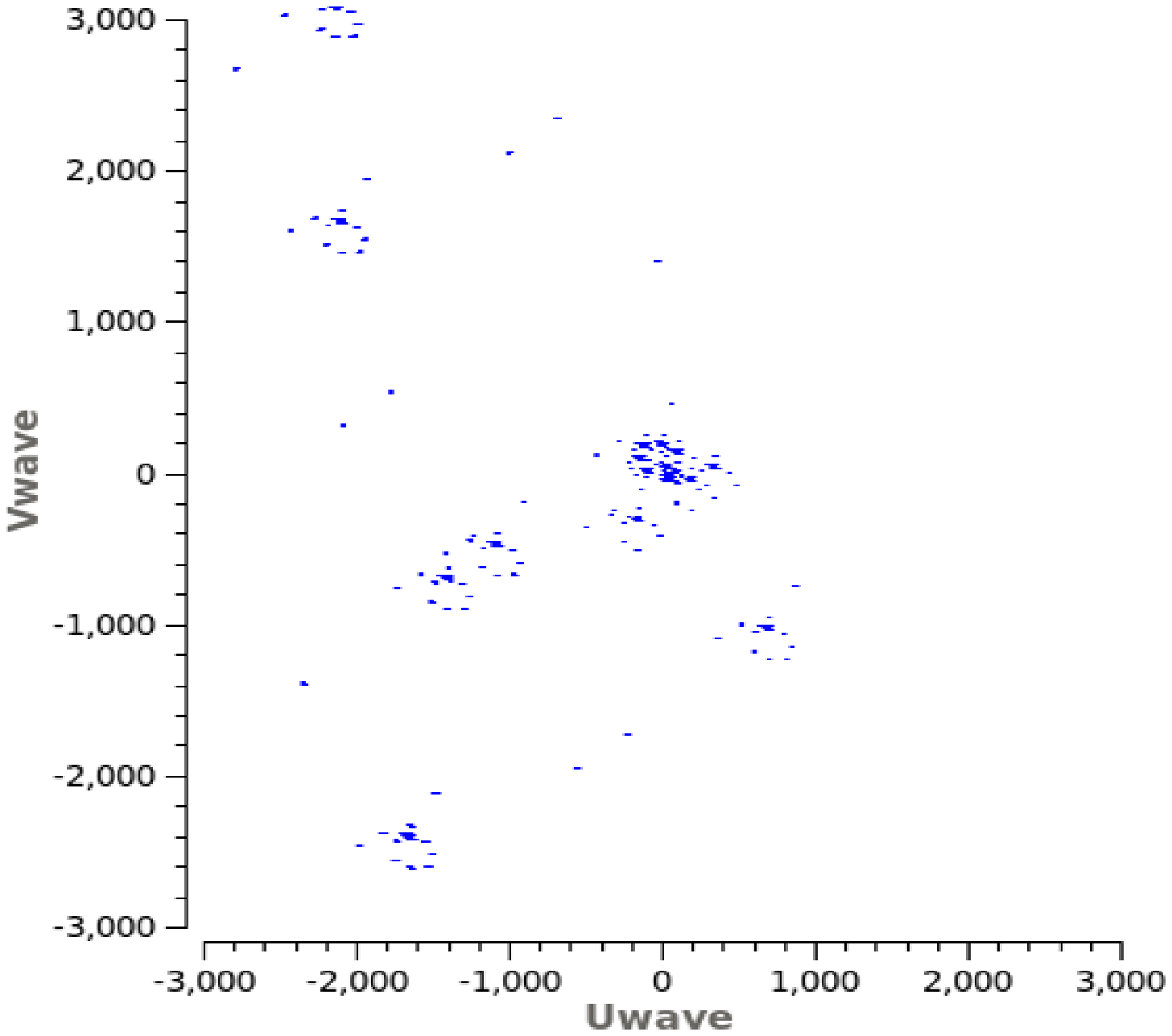}\hspace*{3.2cm}
\includegraphics[width=6cm]{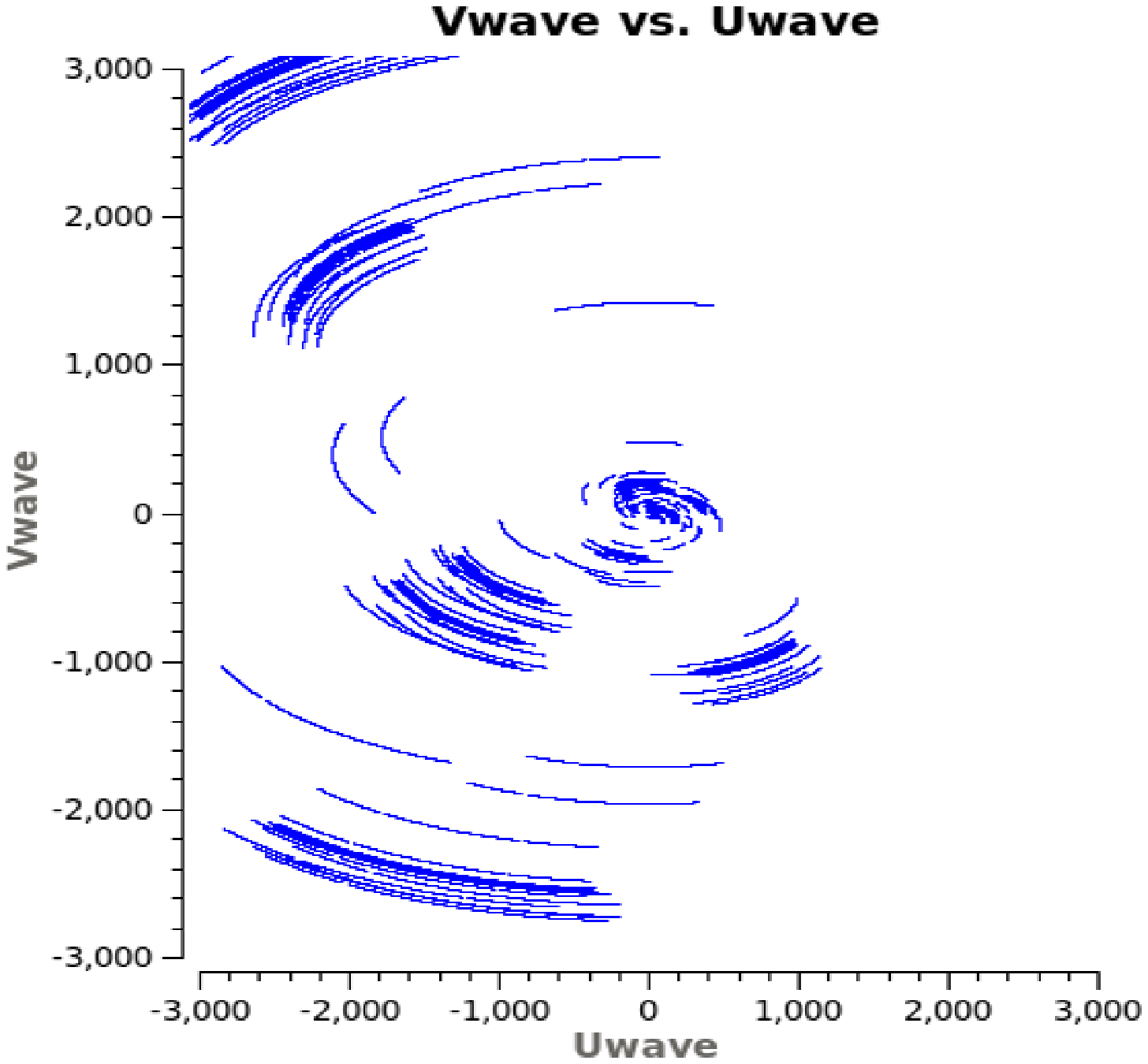}
\caption{uv coverage of the solar observations for snapshot imaging (left) and
  an observing time of three hours (right).}
\label{fig_uv_coverage}
\end{figure*}

Therefore, we use only the first three hours of observing time from
08:02 UT to 11:02 UT. Figure \ref{fig_uv_coverage} shows the uv coverage for
both the usual snapshot solar imaging, and for the three hours used here for
aperture synthesis imaging of the quiet Sun. The uv coverage is improved
substantially.

The Tau A data are calibrated with LOFAR's standard BlackBoard Selfcal (BBS)
system \citep{vanhaarlem13}. The solution was restricted to longer baselines
above 150 wavelengths, since Tau A is a relative compact object with a strong
signal on those baselines, while the Sun is a more extended object with most
intensity in shorter baselines. This approach reduces the risk of solar
interference in the Tau A data used for calibration, and still provides phase
and amplitude corrections for all LOFAR stations.

\subsection{Low-frequency solar images}
After calibration of Tau A, the solutions are transferred to the solar
data. Imaging was performed using the {\em Solar Imaging Pipeline}
\citep{breitling15}, taking the movement of the Sun with respect to the
background sky into consideration. To estimate the quality of the LOFAR
images, it is useful to compare these images with NRH data,
although these are obtained at higher frequencies.

\begin{figure}
\includegraphics[width=8.8cm]{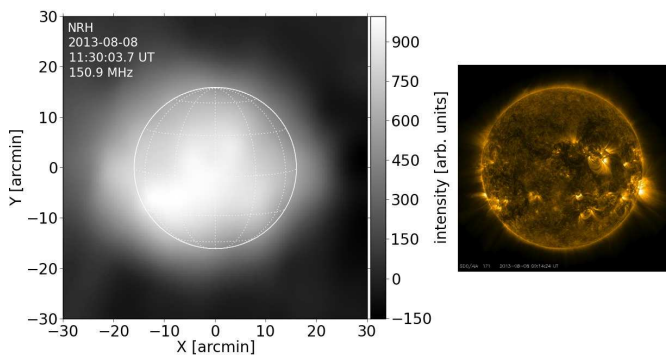}
\caption{Nan\c{c}ay Radioheliograph image at 150.9 MHz (left), and EUV image
  from the {\em Solar Dynamics Observatory} (SDO) in the Fe IX line at
  171\AA{} (right).}
\label{fig_Nancay_SDO}
\end{figure}

Figure \ref{fig_Nancay_SDO} shows images of the Sun from the NRH at their lowest observing frequency of 150.9 MHz, and an EUV
image taken by the {\em Solar Dynamics Observatory} (SDO) in the Fe IX line at
171\AA{}. Both images were taken in the morning hours of 8 August 2013. The
SDO picture has been taken at 9:15 UT, i.e.,~during the LOFAR integration
time. This image shows active regions along the solar equator with some more activity
in the southern hemisphere. The Nan\c{c}ay image is from 11:30 UT; earlier
images had strong artifacts. This image shows the quiet corona with slightly higher
radio intensity south and east of the equator. There was no burst activity
during the LOFAR observing time.

\begin{figure*}[ht]
\includegraphics[width=18cm]{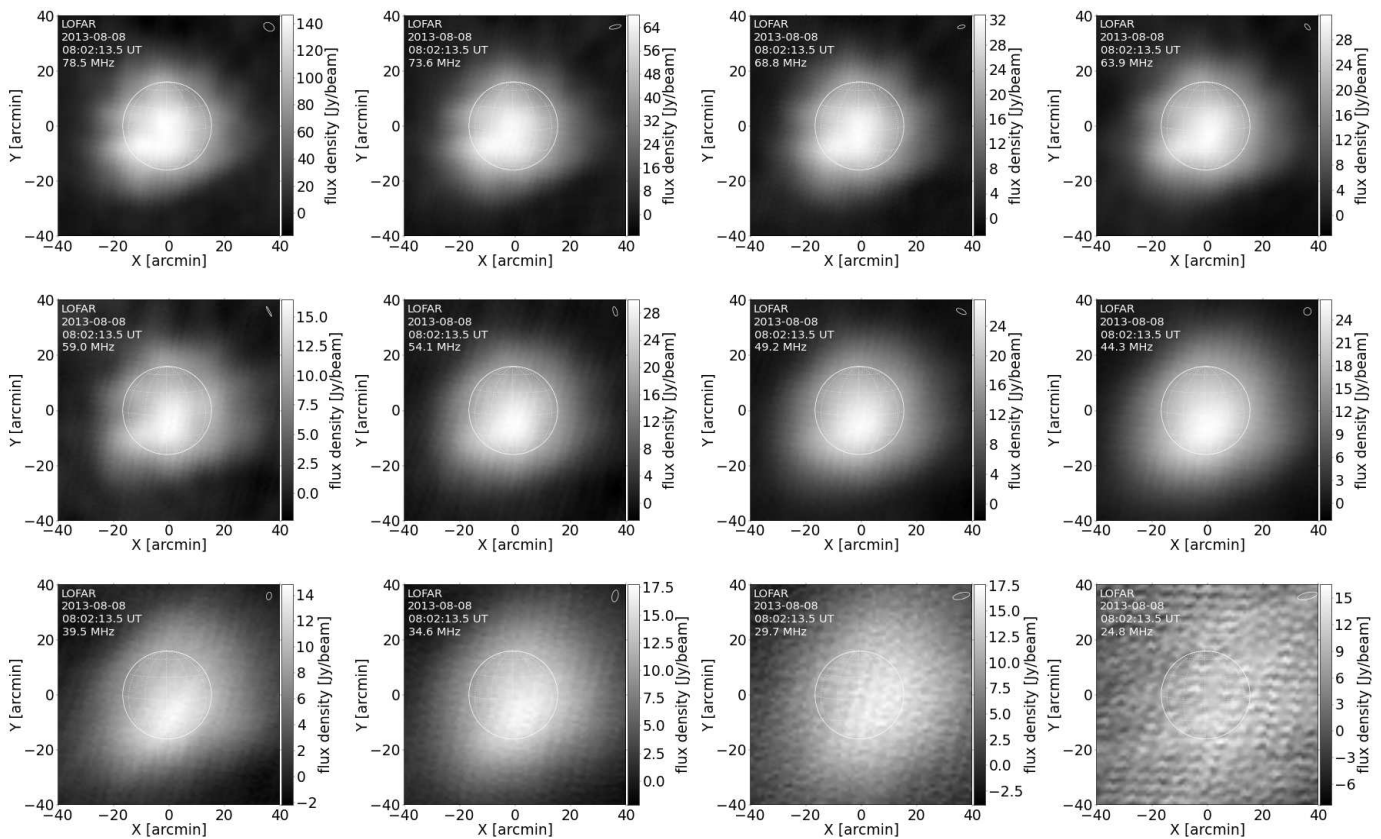}
\caption{Images of the Sun from LOFAR from 8 August 2013, 08:02 -- 11:02 UT, for
  several low-band frequencies ranging from 25 MHz to 79 MHz.}
\label{fig_lofar}
\end{figure*}

Figure \ref{fig_lofar} shows the LOFAR solar images for all observed
frequencies. Imaging was restricted to maximum baseline lengths of 2500
wavelengths, $\lambda$, since longer baselines provide little useful
information from coronal scattering, but could add noise to the images. The
corresponding array beams are indicated by small ellipses in the upper right
corners of each image. Their shapes are frequency dependent because a given
baseline in the LOFAR array, with given length (in km), can correspond to less than $2500\,\lambda$ for lower, and more than $2500\,\lambda$ for higher
frequencies. As the uv coverage is relatively sparse (see
Fig.~\ref{fig_uv_coverage}), such a removal of a baseline from one frequency to
the next can visibly alter the beam shape.

The quiet-Sun image at 79 MHz is in good agreement with the Nan\c{c}ay
image, although the frequency is lower by almost a factor two. Both show the
same structure with slightly enhanced emission toward the southeast of the
disk center. The images change very little for
the next lower frequencies, although the center of brightness moves a bit to
the solar center. Below 50 MHz, it becomes evident that the solar diameter
increases with decreasing frequency. The image quality starts to deteriorate
below 34 MHz. It is still useful at 30 MHz, but very poor for 25 MHz. The lowest frequency of 19 MHz is not shown in Fig.~\ref{fig_lofar} because too
much radio frequency interference (RFI) prevented us from obtaining any useful
images.

There can be many reasons for the lower image quality at lower
frequencies. One reason is an increasing RFI background, another is that
ionospheric influences are not fully corrected when solutions are
transferred from the 50 degree separated Tau A, which is observed through a
different part of the ionosphere. This becomes more severe as the observing
frequency approaches the ionospheric cutoff.

\subsection{Polar and equatorial intensity profiles}
It is evident that the images in Fig.~\ref{fig_lofar} do not show solar disks
with relatively constant intensity and a well-defined limb, but maxima near
the image centers and gradual intensity decrease with increasing angular distance, $\alpha$,
from the centers. Intensity profiles can be derived from the images as a
function of $\alpha$ for each position angle. Averaging over all position angles reduces the image noise and provides smoother profiles, but polar and
equatorial regions should be distinguished.
The SDO image in Fig.~\ref{fig_Nancay_SDO} shows an activity belt around the
solar equator, and a more radial magnetic field configuration around the
poles. Closed magnetic field lines near the equator could lead to a slower
decrease of plasma density with height than at higher latitudes.
\begin{figure}[ht]
\includegraphics[width=8.8cm]{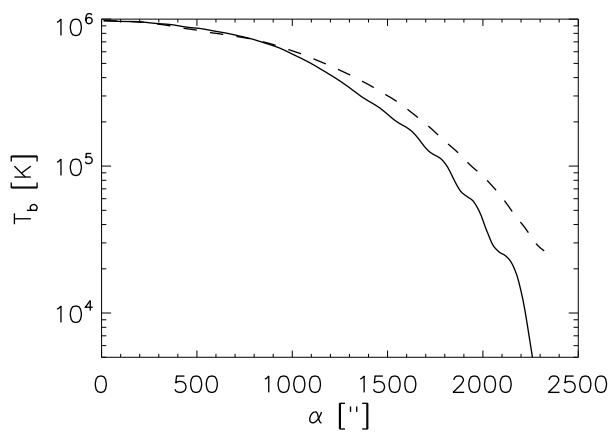}
\caption{Brightness temperature profiles derived from the 74 MHz image for
  polar (solid line) and equatorial (dashed line) regions.}
\label{fig_profile74MHz}
\end{figure}

Figure \ref{fig_profile74MHz} shows the position-angle averaged
intensity profiles within 30 degrees from the polar (north -- south) and
equatorial (east -- west) directions, as
derived from the LOFAR image for 74 MHz. It can be clearly seen that the
intensity indeed falls off more rapidly in the polar direction, which
corresponds to an elongated appearance of the radio Sun in the equatorial direction. This asymmetry is not too pronounced, which is typical for the solar
maximum, but the corona is not radially symmetric.

These results are in agreement with the analysis of \citet{mercier15}, who
studied NRH images of the quiet Sun at frequencies between 150 and 450 MHz, and also found that the radio Sun generally is more
extended in the east-west than in the north-south direction.

The intensity values in Fig.~\ref{fig_profile74MHz} are converted from
Jy/beam to brightness temperatures, using the beam sizes indicated by small
ellipses in Fig.~\ref{fig_lofar} and the Rayleigh-Jeans law. The maximum
brightness temperature near the disk center is close to $10^6$ K. This is significantly higher than the coronal brightness temperatures around 
$7\times10^5$ K usually found by other observers
\citep{mercier15,lantos99,wang87,dulk77}. However, the amplitude calibration
of the LOFAR data is based on solution transfer from the relatively far away
Tau A, as discussed above. Different ionospheric conditions for the solar and calibrator beams could account for this difference.

\section{Radio emission from the quiet Solar corona}
The radio images of the Sun in Fig.~\ref{fig_lofar} appear bigger for lower
frequencies. This is a well-known and expected behavior. In a plasma, radio
waves cannot propagate with a frequency that is less than the local plasma
frequency, $\omega_\mathrm{p}$. The
plasma frequency only depends on the electron density, $N_\mathrm{e}$, and on
no other plasma parameters. In the approximately hydrostatic solar corona, the
density increases toward the Sun. Therefore, a radio wave with frequency $\omega$ can only propagate in the corona, and reach an observer on Earth,
above that coronal height where the electron density, $N_\mathrm{e}$, equals
$\omega^2\,m_\mathrm{e}\,\epsilon_0 / e^2$. In other words, the plasma
frequency defines a surface in the corona above which radio waves can
escape. This radius is denoted as $R_\omega$.

Since the solar corona is highly structured, this surface is generally not a
sphere. The density in, for example,~a coronal hole is lower than in the rest of the
corona, such that a given plasma frequency is found closer to the Sun. However, the
density above active regions is higher, such that the same plasma frequency is
found higher up in the corona. If closed coronal loops with higher density are
present, the surface can be topologically more complex.

Radio emission from this surface is strongly influenced by refraction and optical thickness variation in the corona \citep{smerd50}, which leads to the
observed gradually decreasing intensity profiles in
Fig.~\ref{fig_profile74MHz}.  In the remainder of this paper, we investigate
how such intensity profiles are formed, and  demonstrate how information on
the coronal structure can be derived from these profiles. We only discuss the
density profiles in the polar regions, since they show a simpler,
radially outward magnetic-field geometry for larger $\alpha > 340$'', as
indicated by the EUV image in Fig.~\ref{fig_Nancay_SDO}. Since coronal radio
wave ray paths are refracted away from the solar disk center (see
Fig.~\ref{fig_raytrace_corona} in the next subsection), they traverse only
these radial magnetic fields for $\alpha > 340$''.

A polar, i.e.,~north -- south, profile could be influenced by the equatorial
activity belt for small $\alpha < 340$''. However, the polar and equatorial
intensity profiles in Fig.~\ref{fig_profile74MHz} hardly differ there. Both
show the same slow intensity decrease from the maximum at the disk center,
which continues up to $\alpha = 900$'', where both profiles start
deviating. This is far away from the equatorial activity belt for the polar
profile. Therefore, closed magnetic field structures near the equator should have no
significant influence on the following analysis of polar intensity profiles.

\subsection{Radio wave refraction in the corona}
In the solar corona, the electron cyclotron frequency is usually much smaller
than the plasma frequency. Therefore, the refractive index of a radio wave can
be described by that in a nonmagnetic plasma, $n = (1 - \omega_\mathrm{p}^2 /
\omega^2)^{1/2}$. It approaches zero for $\omega \rightarrow
\omega_\mathrm{p}$, so that radio waves can experience total reflection in the
corona.
This has important consequences for the appearance of the radio Sun as it is
observed from Earth.
\begin{figure}[ht]
\includegraphics[width=8.8cm]{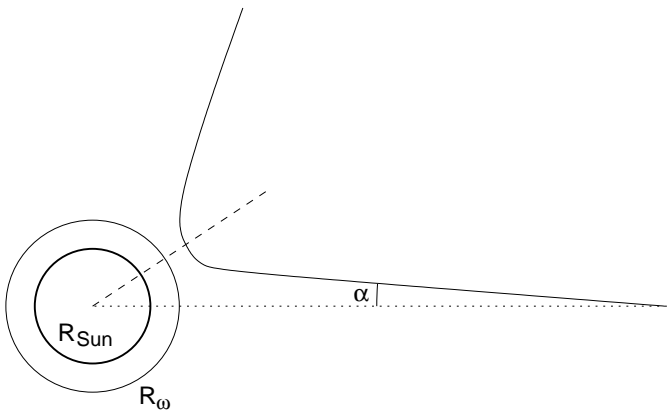}
\caption{Geometry of radio wave refraction in the solar corona.}
\label{fig_raytrace_corona}
\end{figure}

Figure \ref{fig_raytrace_corona} shows the geometry of radio wave
refraction within the solar corona. The observer is located at the far right
and looks into a direction that differs from the solar center by an angle $\alpha$. This idealized picture is based on the assumption of a spherically
symmetric solar corona.

We note that the line of sight can only reach $R_\omega$, i.e.,~the coronal height
where $\omega = \omega_\mathrm{p}$, for observations of the center of the
solar disk. For $\alpha > 0$, the ray
path is deflected at a larger coronal height. Such trajectories of meter waves
through the corona were already discussed by \citet{newkirk61} and
reviewed by \citet{mclean85}.

\subsection{Appearance of the quiet radio Sun to the observer}
The thermal emission from the quiet radio Sun is free-free radiation with emissivity $\epsilon$. The intensity $i(\alpha)$ seen in direction $\alpha$ 
corresponds to its line integral along the ray path for the optically thin
limit. But the solar corona cannot be considered as optically thin, the absorption
of radio waves in the plasma also has to be considered. According to
\citet{mclean85}, it can be calculated as
\begin{equation}
\mu = \frac{\omega_\mathrm{p}^2 \nu_\mathrm{coll}}{\omega^2 c n},
\label{def_mu}
\end{equation}
with collisional frequency $\nu_\mathrm{coll} = 3.6\cdot10^{-6} N_\mathrm{e}
T_\mathrm{e}^{-3/2} (17.7 + 1.5\cdot \ln(T_\mathrm{e}) -
\ln(f))\mathrm{s}^{-1}$ for $N_\mathrm{e}$ in $\mathrm{m}^{-3}$,
$T_\mathrm{e}$ in K, and observing frequency $f = \omega / (2 \pi)$ in Hz. The
absorption length scale can be as short as $0.1\,R_\odot$ for the coronal densities and temperatures discussed in this paper.

Emissivity, $\epsilon$, and absorption, $\mu$, are coupled through
Kirchhoff's Law,
\begin{equation}
\epsilon = \mu \frac{2 k_\mathrm{B} T_\mathrm{e} f^2}{c^2}. \label{def_epsilon}
\end{equation}

When a line integral along a ray path as that in
Fig.~\ref{fig_raytrace_corona} is calculated, the change of the refractive
index along the path also needs to be taken into consideration. The intensity
then evolves according to the radiative transfer equation
\citep[e.g.,][]{mclean85}
\begin{equation}
n^2 \frac{\mathrm{d}}{\mathrm{d}s} \left(\frac{i}{n^2}\right) = \epsilon - 
\mu i. \label{eq_i_int}
\end{equation}

Such a ray-tracing simulation requires a coronal-density model for calculating
not only the emission and absorption, but also the refractive index, which
determines the ray path by means of Snell's law, and enters
Eq.~(\ref{eq_i_int}) directly. As a simple approximation, we use a hydrostatic
density model, which describes the coronal density, or the electron density, as
\begin{equation}
N_\mathrm{e}(r) = N_0 \exp(1 / (H_0 r)), \label{eq_hydrostatic}
\end{equation}
with a parameter
\begin{equation}
H_0 = \frac{k_\mathrm{B} T}{0.6\,m_\mathrm{p}\,g_\odot}
\frac{1}{R^2_\odot} \label{def_H0}
,\end{equation}
which contains the pressure scale height as the first factor, $R_\odot$ is the
solar radius, $g_\odot = 274\,\mathrm{m s}^{-2}$ is the gravitational
acceleration at the coronal base, and 0.6 proton masses is the average
particle mass in the corona \citep{priest82}. The temperature, $T$, is a
scale height temperature determined by both electron and ion temperatures.

$N_\omega = \omega^2 m_\mathrm{e} \epsilon_0 / e^2$ is the density,
where the plasma frequency equals the observing frequency, $\omega$. In other
words, $N_\omega = N_\mathrm{e}(R_\omega)$. Then the density around $R_\omega$
can be written as
\begin{equation}
\frac{N_\mathrm{e}(r)}{N_\omega} = \exp{\left(\frac{1}{H_0}\left(\frac{1}{r} -
\frac{1}{R_\omega}\right)\right)} \label{eq_N_norm}
.\end{equation}
Such a description has the advantage that a global density model for the whole
corona is not required. Only the relative density decrease above $R_\omega$,
with a scale height temperature, $T$, is needed.

Therefore a ray-tracing simulation for given observing frequency, $\omega$,
and angle $\alpha$, depends on two model parameters, $R_\omega$ and a coronal
scale height temperature, $T$. This should be the temperature near $R_\omega$,
since both emission and absorption are strongest at the lowest point of the
ray path through the corona (see Fig.~\ref{fig_raytrace_corona}), which is
near $R_\omega$ and where the density is highest.

Integrating Eq.~(\ref{eq_i_int}) along the ray path is now a two-step
process. At first, the ray path needs to be determined, starting at the
position of the observer 1 AU away from the Sun, and proceeding into direction $\alpha$. The density model, Eq.~(\ref{eq_N_norm}), provides the refractive
index, which leads to the ray path by Snell's Law, all
the way toward the Sun, through the corona, and back into interplanetary space. The ray path is calculated for a solar distance up to 1 AU.

The second step is integrating equation (\ref{eq_i_int}) back
from interplanetary space beyond the Sun, through the corona, toward the observer. The initial condition is $i = 0$, i.e.,~it is assumed that no
emission is coming from beyond that point. Since the emissivity scales with
$N_\mathrm{e}^2$, and the solar wind density at 1 AU is at least by a factor
of $10^{-8}$ lower than that in the lower corona, the arbitrary choice of a
starting point of 1 AU behind the Sun is not critical.

\begin{figure}[ht]
\includegraphics[width=8.8cm]{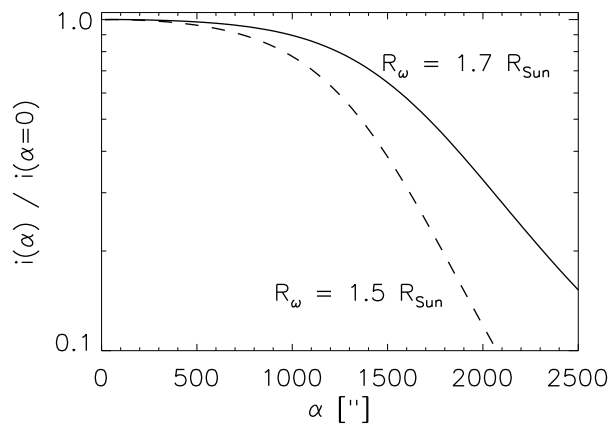}
\caption{Intensity profiles from two ray-tracing simulations with $R_\omega =
  1.7\,R_\odot$ (solid line) and $R_\omega = 1.5\,R_\odot$ (dashed line), both
  for a coronal scale height temperature of $T = 1.4\times 10^6\,\mathrm{K}$
  and an observing frequency $f = 54\,\mathrm{MHz}$.}
\label{fig_i_profile_Romega}
\end{figure}
Figure \ref{fig_i_profile_Romega} shows the results of such simulations for
two different values of $R_\omega$. The other simulation parameters, i.e.,~a
coronal scale height temperature $T = 1.4\times 10^6\,\mathrm{K}$
and an observing frequency $f = 54\,\mathrm{MHz}$, are identical. The electron
temperature, $T_\mathrm{e}$, which enters Eq.~(\ref{eq_i_int}), is set to be
equal to the scale height temperature, $T$.

This is just the simplest approach to such a model. In general, $T$ and
$T_\mathrm{e}$ can differ, for example, \citet{mercier15} deduced a much lower
$T_\mathrm{e} = 6.2\times10^5$ K. In the optically thick limit, 
i.e.,~small $\alpha$ near the disk center, $T_\mathrm{e}$ determines the
brightness temperature. In the optically thin limit, i.e.,~large $\alpha$
traversing the outer corona, the intensity, Eq.~(\ref{eq_i_int}), is just a
line integral over the emissivity $\epsilon$, Eq.~(\ref{def_epsilon}), which
contains a factor $T_\mathrm{e}^{-1/2}$. In this case, the slope of an
intensity curve in a semi-logarithmic plot such as
Fig.~\ref{fig_i_profile_Romega} is independent of $T_\mathrm{e}$. For
$R_\omega = 1.5\,R_\odot$ and $\alpha = 1500$'', the optical depth $\tau =
0.2$. For $\alpha = 2000$'', it has decreased to $\tau = 0.05$.
Thus the choice of $T_\mathrm{e}$ is not critical for the larger $\alpha$
in Fig.~\ref{fig_i_profile_Romega}, where the intensity profiles show a strong
$R_\omega$ dependence.

It follows from Fig.~\ref{fig_i_profile_Romega} that the radial extent of the
radio Sun increases with higher $R_\omega$. This is not surprising, since a
higher $R_\omega$ just means that the sphere around the Sun, within which radio waves with frequency $\omega$ cannot propagate, is larger.

A radius of $1.5\,R_\odot$ appears under an angle $\alpha = 1350$'' to an
observer on Earth. But the intensity values fall already for much smaller $\alpha$ below their peak value at the solar disk center. The same behaviors
 as those in
Fig.~\ref{fig_profile74MHz} are found in observed intensity profiles. The reason for this lies in the ray path, as depicted in Fig.~\ref{fig_raytrace_corona}.
For increasing $\alpha$, the turning point of the ray path tends to be at a
larger coronal height, where the plasma density and therefore the emissivity
is lower.

\begin{figure}[ht]
\includegraphics[width=8.8cm]{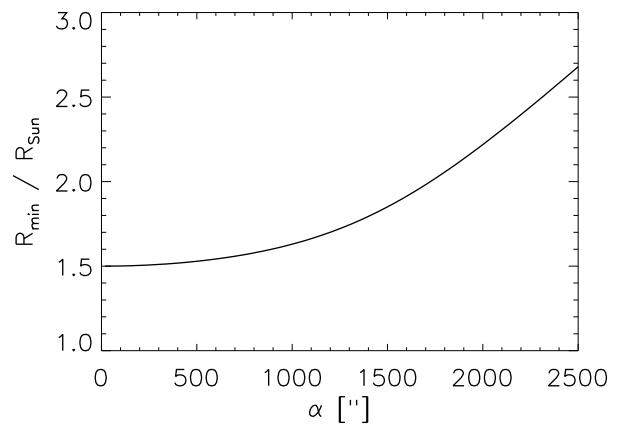}
\caption{Height of the closest point of the ray path to the Sun, as a function
  of viewing angle $\alpha$, for $R_\omega = 1.5\,R_\odot$, a scale height
  temperature $T = 1.4\times 10^6\,\mathrm{K}$, and observing frequency $f =
  54\,\mathrm{MHz}$.}
\label{fig_rmin}
\end{figure}

The plot of this closest distance, $R_\mathrm{min}$, as a function of $\alpha$
in Figure \ref{fig_rmin} confirms this conjecture. For higher $\alpha$, the
ray path probes only higher parts of the corona. For $\alpha = 1350$'',
$R_\mathrm{min}$ has increased to $1.7\,R_\odot$ from the original
$1.5\,R_\odot$ for $\alpha = 0$. For a coronal temperature of
$1.4\times 10^6\,\mathrm{K}$ this corresponds to 0.8 pressure scale
heights at altitudes around $1.6\,R_\odot$. Since the emissivity of quiet-Sun
radio emission scales with $N_\mathrm{e}^2$, and the corona is not optically
thick for these radio waves, this explains why both observed
(Fig.~\ref{fig_profile74MHz}) and model (Fig.~\ref{fig_i_profile_Romega})
intensity profiles decrease for relatively small $\alpha$.

\begin{figure}[ht]
\includegraphics[width=8.8cm]{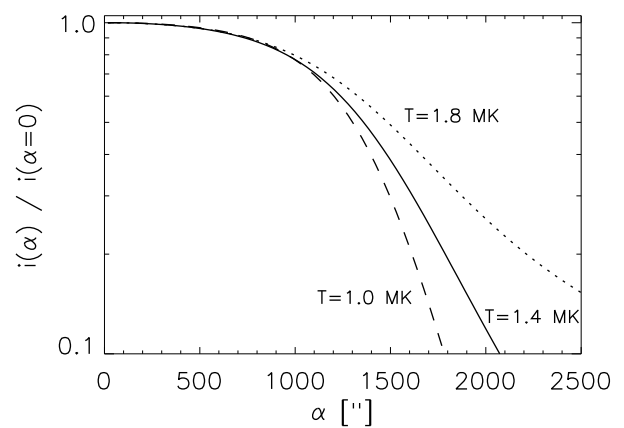}
\caption{Intensity profiles from three ray-tracing simulations with $R_\omega =
  1.5\, R_\odot$ and different scale height temperatures: $T =
  1.4\times 10^6\,\mathrm{K}$ (solid line), $T = 1.0\times 10^6\,\mathrm{K}$
  (dashed line), and $T = 1.8\times 10^6\,\mathrm{K}$ (dotted line), all
  for an observing frequency $f = 54\,\mathrm{MHz}$.}
\label{fig_i_profile_Tcor}
\end{figure}
The ray-tracing simulations depend not only on $R_\omega$, but also on the
coronal temperature. Intensity profiles for three different $T$ are
plotted in Figure \ref{fig_i_profile_Tcor}. Generally, the higher the scale
height temperature, the further out the radio Sun extends. The slope of the
intensity profile becomes temperature dependent for higher $\alpha > 1200$'',
it decreases with increasing $T$. But, unlike the variation of
$R_\omega$ in Fig.~\ref{fig_i_profile_Romega}, there is little influence of
$T$ on the intensity profile for lower $\alpha$.

The reason for this also lies in the $\alpha$  dependence of the ray-path
geometry. For higher $\alpha$, the ray path traverses only higher parts of the
corona (see Fig.~\ref{fig_rmin}). It follows from Eq.~(\ref{eq_N_norm}) that
further away from $R_\omega$, the pressure scale height, and thus
$T$, has an increasing impact on the plasma density there, and thus
on both the free-free radio wave emission/absorption and the actual
ray path.

The discussion of Figures \ref{fig_i_profile_Romega} and
\ref{fig_i_profile_Tcor} leads to the conclusion that the ray-tracing
simulation parameters $R_\omega$ and $T$ influence the resulting
intensity profiles across the radio Sun in distinct ways. Therefore, it is
possible to fit both of these to an observed intensity profile. Since the
ray-tracing profiles coincide for smaller $\alpha < 500$'', it is the larger
$\alpha$ that determine the fitting parameters. This has the consequence that
the slightly enhanced emission south and, especially for the higher
frequencies, east of the disk center found in Fig.~\ref{fig_lofar} does not
influence the results.

\subsection{Coronal density model based on LOFAR images}
Before an observed intensity profile can be fitted with ray-tracing
simulations with variable values of $R_\omega$ and $T$, the finite
angular resolution of radio observations has to be considered. To take this
into account, the model intensity profiles are convolved with a Gaussiam beam
profile. For the observations presented in this work, a FWHM beam of 150'' is
typical. However, this has little effect on the simulated profiles.

\begin{figure}[ht]
\includegraphics[width=8.8cm]{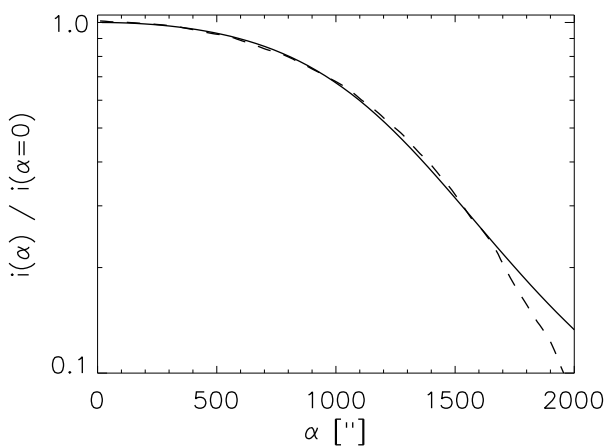}
\caption{Model (solid line) and observed (dashed line) intensity profiles for
  an observing frequency $f = 54\,\mathrm{MHz}$. The simulation parameters are
  $R_\omega = 1.36\,R_\odot$ and $T = 1.85\times 10^6\mathrm{K}$.}
\label{fig_fit_54MHz}
\end{figure}

Figure \ref{fig_fit_54MHz} shows the result of such a fit to the observed 54
MHz polar intensity profile. As stated above,the polar regions of the
Sun show a simple, radial magnetic field geometry that matches well the
ray-tracing simulations with their local hydrostatic density model in the radial direction.
The simulation parameters that provide the best fit are $R_\omega
= 1.36\,R_\odot$, and a scale height temperature of $T = 1.85\times
10^6\,\mathrm{K}$. This is a surprisingly high value, much more than the usual $T
= 1.0 - 1.4\times 10^6\,\mathrm{K}$ in coronal density models such as those of
\citet{mann99} or \citet{newkirk61}.

But this high temperature does provide the best fit; the model intensity
profile is in close agreement with the observed profile for $\alpha < 1600$''.
A variation of $T$ mainly leads to different slopes of the
simulated profile for larger $\alpha > 1000$'' (see
Fig.~\ref{fig_i_profile_Tcor}). So a lower temperature value would lead to a
steeper intensity profile, not in agreement with the observed profile. For larger
$\alpha > 1600$'', the slope of the observed intensity profile starts to be
steeper than the model profile, as if a lower temperature value was relevant
there.

The scale height temperature value $T$ has its strongest impact on the outer,
high $\alpha$ part of simulated intensity profiles. This enables a quick test
of the method of line-of-sight integration over emissivities and absorption,
which is based on Eqs.~(\ref{def_mu})--(\ref{eq_i_int}). If the emission
and absorption terms are replaced by a simple line integral over
$N_\mathrm{e}^2$,
i.e.,~all effects of finite optical depth are ignored, then it is found that
this manipulation has little effect on the slope of the outer part of the profile. The reason for this behavior is that for higher $\alpha$, the ray
path only traverses higher, and therefore more dilute regions, of the corona,
where the optical depth is low. But still, the same high coronal temperature
is needed to fit the observation. This test indicates that the high
temperature is a robust result, and not an artifact based on a flaw of the
ray-tracing simulations using Eqs.~(\ref{def_mu})--(\ref{eq_i_int}).

In order to study the dependence of the fit quality on the model parameters
$R_\omega$ and $T$, and to estimate error margins for these, we define
residuals $\chi^2$ as a measure for the difference between model and observed
intensity profiles in a semi-logarithmic plot, that is,
\begin{equation}
\chi^2 = \int_0^{1600\mathrm{''}} \left(\ln i_\mathrm{model}(\alpha) -
\ln i_\mathrm{observed}(\alpha)\right)^2\,\mathrm{d}\alpha.
\label{eq_chi2}
\end{equation}
Minimizing $\chi^2$ corresponds to a least-square fit to the observed
profile. The upper boundary of the integral was chosen as $\alpha =1600$'',
where both profiles start deviating from each other.
\begin{figure}[ht]
\includegraphics[width=8.8cm]{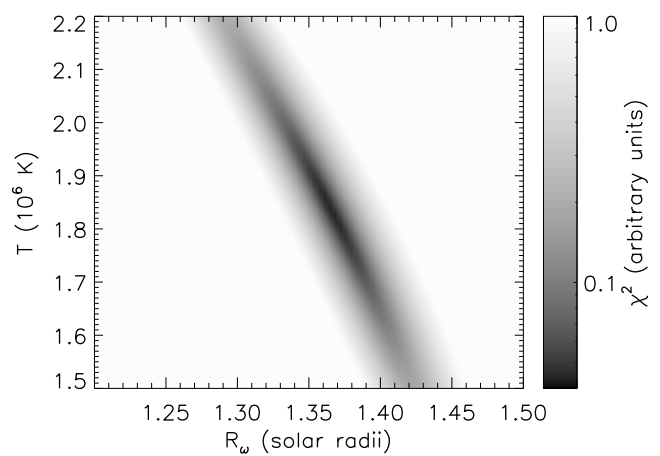}
\caption{Residuals $\chi^2$ between model and observed intensity profiles, as
  functions of fitting parameters $R_\omega$ and $T$.}
\label{fig_fit_quality_54MHz}
\end{figure}
Fig.~\ref{fig_fit_quality_54MHz} shows the residuals $\chi$ as functions of
$R_\omega$ and $T$. Low values of $\chi^2$ form a line in $R_\omega$ - $T$
space, with a minimum at $R_\omega = 1.36\,R_\odot$ and $T = 1.85\times
10^6\mathrm{K}$, which provide the best fit and minimize $\chi^2$. The width of
the area with lowest $\chi^2$ provides estimates for error margins of $\pm
0.01\,R_\odot$ for $R_\omega$ and $\pm 5\times 10^4$ K for $T$.
Significantly lower temperatures, for example,~$T = 1.4\times 10^6$ K, increase the residuals $\chi^2$ by more than one order of magnitude.

\begin{figure}
\includegraphics[width=8.8cm]{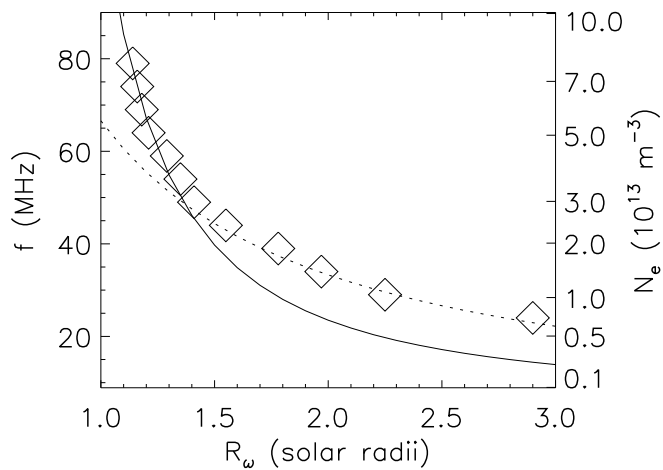}
\caption{Observed frequencies, $f$, as a function of fitted $R_\omega$,
  together with a hydrostatic density model (solid line) and $1/r^2$ density
  model (dotted line).}
\label{fig_densitymodel}
\end{figure}
The same fitting procedure is now repeated for all other observed
frequencies. Figure \ref{fig_densitymodel} shows the frequencies, $f$, as a function of the fitted values for $R_\omega$. Since $R_\omega$ is where the
local plasma frequency equals the observation frequency, this plot corresponds
to a density profile, $N_\mathrm{e}(r)$. The solid line represents a
hydrostatic coronal density model based on a constant coronal temperature of
$T = 2.2\times 10^6$ K and
a density $N_0 = 1.6\times 10^{14}\,\mathrm{m}^{-3}$ at the coronal base. The
dotted line delineates a density profile that scales as
$N_\mathrm{e}(r)\propto 1/r^2$, which is discussed below.
\begin{figure}
\includegraphics[width=8.8cm]{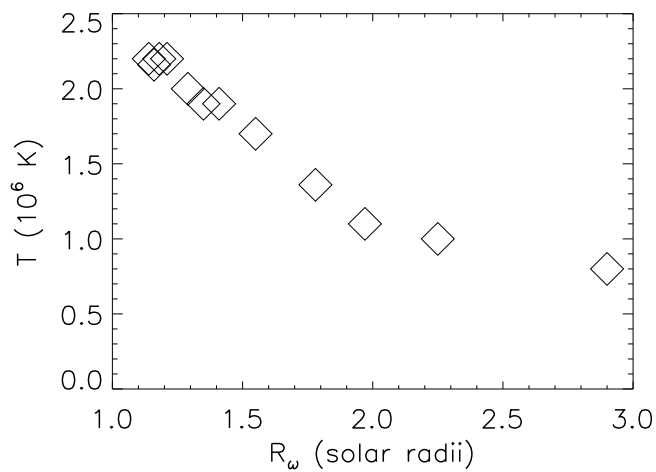}
\caption{Fitted scale height temperature values, as a function of the
  $R_\omega$, fitted to each observed frequency.}
\label{fig_Tmodel}
\end{figure}
Figure \ref{fig_Tmodel} shows the corresponding coronal temperatures. It is
obvious that the $T = 1.9\times 10^6$ K found for $f = 54$ MHz is not even the
highest value. In the low corona, i.e.,~for the highest frequencies, $T$
reaches up to $2.2\times 10^6$ K. We note that these temperature values were
obtained for each observed frequency independently.

The temperature $T = 2.2\times 10^6$ K that was used in the hydrostatic
model fit of the density values in the low corona (see
Fig.~\ref{fig_densitymodel}) provides the best fit to the data. A lower temperature leads to a density decrease with
height that is too fast. This result is independent of the fitting procedure for individual
intensity profiles as in Fig.~\ref{fig_fit_54MHz}. So the high scale height
temperature in the low corona is a consistent result that has been obtained
from two independent methods.

The coronal temperature profile in Fig.~\ref{fig_Tmodel} also provides an explanation for the deviation of the fit in Fig.~\ref{fig_fit_54MHz} for
higher viewing angles $\alpha > 1600$''. With increasing $\alpha$, the ray
path traverses higher layers of the corona, where the temperature is indeed
lower, as already indicated in the discussion of the fit in
Fig.~\ref{fig_fit_54MHz}. The poor fit for higher $\alpha$ is therefore an
artifact of the model assumption of a constant scale height temperature. For
smaller $\alpha$, this is not critical, since the minimum distance of the ray
path to the Sun depends less on $\alpha$ (see Fig.~\ref{fig_rmin}).

\section{Discussion}
It has been shown that LOFAR images of the quiet solar corona allow for the
derivation of coronal density and temperature values. This is performed individually for each
observed frequency by fitting a model intensity profile across
the radio Sun, based on a ray-tracing simulation, to the observed profile.
The density is obtained in the form of the radius $R_\omega$, where the local
plasma frequency equals the observed frequency, $\omega$. In this paper, we
only studied intensity profiles in the polar direction, with their
simpler, radial magnetic field geometry.

The results show surprisingly high scale height temperatures of $2.2\times
10^6$ K in the low corona. This temperature describes the density decrease
with altitude, that is directly observed in coronal white-light
scattering. Density models based on such observations lead to much lower
temperatures of $T =1.4\times 10^6$ K, such as those of
\citet{newkirk61}. Other coronal models, such as that of \citet{mann99}, who
fits a Parker-type solar coronal and wind model to 
observations ranging from the corona up to 1 AU, also indicate temperatures
closer to $T = 10^6$ K.

But if the $R_\omega$ and temperatures from all
observed frequencies are combined, they lead to a coherent picture. The
temperature profile (see Fig.~\ref{fig_Tmodel}) shows a continuous decrease
from the $2.2\times 10^6$ K in the low corona to less than $10^6$K in the
outer corona. The $R_\omega$ values can be combined into a density profile,
which can 
be fitted by a hydrostatic model for small heights below $1.4\,R_\odot$. But
the hydrostatic model has to be based on the same high temperature of $T =
2.2\times 10^6$ K. It should be noted that the scale height temperature in
such a single-fluid model is a combination of electron and ion temperatures.

This agreement is an indication that the temperature results are correct, since
the fits to the intensity profiles were carried out for each frequency, i.e.,~coronal
height, independently. The density model that results from all values for
$R_\omega$ combined together, thus is consistent with such a high temperature.
The reason for this is unclear, since the SDO images indicate no signs of activity in the polar regions. 

The hydrostatic model is based on a density at the coronal base of $N_0 =1.6\times 10^{14}\,\mathrm{m}^{-3}$. This is very low for the quiet Sun, it
corresponds to just one-fifth of the value in the model of \citet{newkirk61}.
It would be more typical for a coronal hole, although a coronal hole is
not visible in the SDO images.

The high temperatures of up to $T = 2.2\times 10^6$ K have been found at low
radial distances of less than $1.5\,R_\odot$. This altitude range has been
studied by \citet{mercier15}, at higher frequencies with the NRH. They have also found relatively high scale height
temperatures in the north--south direction in their Fig.~6, varying between
$1.65\times 10^6$ K and $2.24\times 10^6$ K. So there is some consistency with these observations.

Extreme UV observations are another source for coronal temperature data, i.e.,~coronal
electron temperatures, $T_\mathrm{e}$. However, it is a long-standing issue
that the observed $T_\mathrm{e}$ on the order of $10^6$ K are higher than brightness temperatures at the solar disk center of about $7\times 10^5$ K
\citep{dulk77}. Reconciling these temperatures requires further model assumptions
\citep{chiuderi_drago99}. The scale height temperatures, $T$, discussed here
is not $T_\mathrm{e}$, but results as an average of both electron and ion temperatures. Even a $T_\mathrm{e} = 10^6$ K would require an average ion
temperature of about $T_\mathrm{i} = 3.4\times 10^6$ K for a scale height
temperature of $2.2\times 10^6$ K. Minor ions, such as O${}^{5+}$, are observed
to have very high temperatures up to $10^8$ K \citep{antonucci12}, but these
values are found higher up in the corona above $2\,R_\odot$ and not for the
bulk of protons. Below $2\,R_\odot$, O${}^{5+}$ temperatures drop to $8\times
10^6$ K at $1.4\,R_\odot$, so minor ions with their relative abundances well
below $10^{-3}$ \citep{grevesse98} should not have a significant influence on
lower coronal scale heights.
The issue of high scale height temperatures in the low
corona in the radio data has to be left open for future research.

For larger coronal heights, around $1.5\,R_\odot$, the $R_\omega$ become
larger than predicted by the hydrostatic model. In other words, the density
falls off slower than hydrostatically. The temperature is already lower in 
this region, which should lead to an even faster density decrease than
predicted by the $T = 2.2\times 10^6$ K hydrostatic model.

This is the expected behavior for the transition from the subsonic plasma
flow in the corona into the supersonic solar wind, where the density falls off
as $r^{-2}$ once a constant solar wind speed has been reached. A significant
deviation from a hydrostatic profile is to be expected once the sonic Mach
number, $M$, does not satisfy the condition $M \ll 1$ anymore. The dotted line
in Fig.~\ref{fig_densitymodel} shows such an $N_\mathrm{e}(r)\propto 1/r^2$
profile. This profile fits the data well in the upper corona, even as low as
$1.5\,R_\odot$.

This indicates a transition into the supersonic solar wind that is located
rather low in the corona, but still in agreement with some solar wind models
\citep{axford99}. Furthermore, similar nonhydrostatic density profiles have
also been found in the analysis of type III radio bursts \citep{cairns09}.

The ray tracing simulations, based on Eqs.~(\ref{def_mu})--(\ref{eq_i_int})
used to calculate model intensity profiles, which are then fitted to 
the observed profiles as in Fig.~\ref{fig_fit_54MHz}, stem from an
important assumption. The profiles are calculated as if the coronal density
decreases hydrostatically above $R_\omega$. But in the transition to the solar
wind, the corona is not hydrostatic anymore. However, this hydrostatic
assumption is not used to calculate a density model of the whole
corona. Instead, it only describes the density decrease just above
$R_\omega$. At larger heights, the free-free radiation emissivity decreases
with the square of the density, such that there is only a minor contribution
to the line-of-sight integrals the model intensity profiles are based on.

To the first order, the density decrease above $R_\omega$ can be described by
the hydrostatic model, Eq.~(\ref{eq_N_norm}), even if the corona is not
hydrostatic. The derivative of Eq.~(\ref{eq_N_norm}) in $r$ at the position $r
= R_\omega$ can be written as
\begin{equation}
\left.\frac{\mathrm{d} \ln N_\mathrm{e}(r)}{\mathrm{d} r}\right|_{r = R_\omega} =
-\frac{1}{H_0 R_\omega^2}. \label{eq_N_gradient}
\end{equation}
Replacing the left-hand side of Eq.~(\ref{eq_N_gradient}) by the logarithmic
coronal density gradient, which has been derived from the data, then allows for
calculating $H_0$, and through Eq.~(\ref{def_H0}) the scale height
temperature $T$. For a nonhydrostatic corona, this is not the actual
temperature, but just a parameter describing the density decrease with
height. If the coronal density, for example,~falls off slower than hydrostatically,
this $T$ would be higher than the temperature there.

For the temperature values shown in Fig.~\ref{fig_Tmodel}, this has the
consequence that those from coronal heights above $1.5\,R_\odot$ should be
treated with caution. But in the low corona, where the high $T = 2.2\times
10^6$ K have been found, a hydrostatic density model seems to be
applicable. So the hydrostatic model assumption in the ray-tracing simulations
seems unlikely to be responsible for the high temperatures found there.

\section{Summary and outlook}
The analysis of these first quiet-Sun Earth-rotation interferometric imaging
observations with LOFAR show their potential for investigating the coronal
structure. Density models can be derived and the transition from the
hydrostatic lower solar corona into the supersonic solar wind can be studied.

But the interpretation of coronal radio images is not straightforward. The
effects of the refractive medium of the solar corona on radio wave
propagation, including free-free wave emission and absorption, need to be taken into account. Considering these effects, it is possible to calculate
intensity profiles across the Sun with ray-tracing simulations.

For the first results presented here, we focused on the simple, radial
magnetic-field geometry above the poles of  the quiet Sun. More complex structures
such as closed loops were not studied.

The Sun was only observed on a few selected frequencies, with 5 MHz gaps
between them, for these early LOFAR data. Future observations provide a
continuous frequency coverage in LOFAR's low band of 10 -- 90 MHz, with a frequency resolution that corresponds to the LOFAR 195 kHz sub-band width.

With continuous frequency coverage, it will be possible to derive
iteratively a single coronal density model that can be used for all ray-tracing
simulations for different frequencies. This eliminates the need for a local
hydrostatic density model in the data analysis, and thereby the need to fit
model intensity profiles not only to $R_\omega$, but also to a scale height
temperature value.
An example would be studies of equatorial density profiles that could
traverse closed magnetic field lines connecting active regions such as those in
the SDO image (Fig.~\ref{fig_Nancay_SDO}). For such structures, the model
simplification of local hydrostatic equilibrium is at least questionable. This
is why we did not perform the analysis in the present paper.

Scale height temperatures still can be derived from the
barometric scale height of the density profile as long as the corona can be
considered in hydrostatic equilibrium. This will provide a test of whether the
high temperatures of more than $2\times 10^6$ K will still be found.

Such an improved method will lead to a better analysis of data for the upper
corona, where the transition into the supersonic solar wind occurs and where
a local hydrostatic model is a rough approximation. Furthermore, it will
enable studies of more complex coronal structures, where a local hydrostatic
density model is not applicable at all. Thus not only the quiet Sun and coronal
holes, but also closed loops and streamer-like structures will be
investigated.

\begin{acknowledgements}
C.~Vocks, G.~Mann, and F.~Breitling acknowledge the financial support by the
German Federal Ministry of Education and Research (BMBF) within the framework
of the project {\em D-LOFAR} (05A11BAA) of the Verbundforschung.

This paper is based on data obtained with the International LOFAR Telescope
(ILT). LOFAR (van Haarlem et al. 2013) is the Low Frequency Array designed and
constructed by ASTRON. It has facilities in several countries, which are owned
by various parties (each with their own funding sources) and which are
collectively operated by the ILT foundation under a joint scientific policy.

The authors wish to thank the referee for constructive comments that have helped to improve the manuscript significantly.
\end{acknowledgements}

\bibliographystyle{aa}
\bibliography{vocks_qS}
\end{document}